\documentclass[12pt]{article}

\usepackage{times}

\usepackage{epsfig}
\usepackage{graphicx}
\usepackage{amsmath,amssymb}

\newcommand{\beq}[1]{
\begin{equation}
\label{e#1} }

\newcommand{\eeq}{
\end{equation}
}

\newenvironment{sciabstract}{%
\begin{quote} \bf}
{\end{quote}}

\newcounter{lastnote}

\title{THz electrical writing speed in an antiferromagnetic memory}
\author
{K.~Olejn\'{\i}k,$^{1,\#}$ T.~Seifert,$^{2}$ Z.~Ka\v{s}par,$^{1,3}$ V.~Nov\'ak,$^{1}$ P.~Wadley,$^{4}$ R.~P.~Campion,$^{4}$ \\ M.~Baumgartner,$^{5}$ P.~Gambardella,$^{5}$ P.~N\v{e}mec,$^{3}$ J.~Wunderlich,$^{1,6}$  \\ J. Sinova,$^{1,7}$  M.~M\"uller,$^{2}$ T.~Kampfrath,$^{2}$ T.~Jungwirth$^{1,4}$\\
\\
\normalsize{$^{1}$Institute of Physics, Academy of Sciences of the Czech Republic,}\\ 
\normalsize{Cukrovarnick\'a 10, 162 00 Praha 6, Czech Republic}\\
\normalsize{$^{1}$Fritz Haber Institute of the Max Planck Society, 14195 Berlin, Germany}\\
\normalsize{$^{3}$Faculty of Mathematics and Physics, Charles University,}\\ 
\normalsize{Ke Karlovu 3, 121 16 Prague 2, Czech Republic}\\
\normalsize{$^{4}$School of Physics and
Astronomy, University of Nottingham,}\\
\normalsize{Nottingham NG7 2RD, United Kingdom}\\
\normalsize{$^{5}$Department of Materials, ETH Z\"urich, H\"onggerbergring 64,}\\
\normalsize{CH-8093 Z\"urich, Switzerland}\\
\normalsize{$^{6}$Hitachi Cambridge Laboratory, J. J. Thomson Avenue,}\\ 
\normalsize{Cambridge CB3 0HE, United Kingdom}\\
\normalsize{$^{7}$Institut f\"ur Physik, Johannes Gutenberg Universit\"at Mainz,}\\ 
\normalsize{55128 Mainz, Germany}\\
\\
\normalsize{$^\#$To whom correspondence should be addressed; E-mail:  olejnik@fzu.cz.}
}

\begin{document}
\baselineskip24pt

\maketitle 

\begin{sciabstract}

The speed of writing of state-of-the-art ferromagnetic memories is physically limited by an intrinsic GHz threshold.  
Recently, an alternative research direction has been initiated by realizing memory devices based on antiferromagnets
in which spin directions periodically alternate from one atomic lattice site to the next. In our work we experimentally demonstrate at room temperature that the speed of reversible electrical writing in a memory device can be scaled up to THz using an antiferromagnet. Efficient current-induced spin-torque mechanism is responsible for the switching in our memory devices throughout the twelve orders of magnitude range of writing speeds from Hz to THz. Our work opens the path towards the development of memory-logic technology reaching the elusive THz band.

\end{sciabstract}

Magnetic random access memories\cite{Chappert2007,Brataas2012,Kent2015} represent the most advanced example of spintronic devices that are foreseen to become the leading  alternatives to CMOS for the "beyond Moore's law" information technologies\cite{Waldrop2016}. Among the unresolved fundamental problems in spintronics is the electrical writing speed. Whether using Oersted fields or advanced spin-torque mechanisms\cite{Ralph2008,Chernyshov2009,Miron2011b,Liu2012,Brataas2012,Bedau2010,Garello2014}, the writing speed has a physical limit in ferromagnetic memories in the GHz range beyond which it becomes prohibitively energy-costly\cite{Garello2014,Prenat2016}. The interest  in antiferromagnetic memories is driven by the vision of ultra-fast operation far exceeding the GHz range \cite{Marrows2016}. Recently, THz writing speed has become a realistic prospect with  the experimental discovery\cite{Wadley2016} of the electrical switching in the CuMnAs antiferromagnet by a staggered spin-torque field\cite{Zelezny2014}  at ambient conditions. This was followed by upscaling of the writing speed  to the GHz range, and by demonstrating a fabrication compatibility with III-V semiconductors or Si, and a device compatibility with common microelectronic circuitry\cite{Olejnik2017}. 

These initial experiments verified several unique features of antiferromagnetic bit cells, including their magnetic field hardness, absence of fringing stray fields, and neuron-like multi-level memory-logic functionality\cite{Wadley2016,Olejnik2017,Jungwirth2016}.  The results have been already replicated in another suitable antiferromagnet Mn$_2$Au\cite{Bodnar2017,Meinert2017}. However, the envisaged THz electrical writing speed\cite{Zelezny2014,Jungwirth2016,Roy2016,Marrows2016}  in antiferromagnetic memories has not been experimentally demonstrated prior to this work. 

{\bf Switching mechanism and measurement setup.} The principle of the current-driven antiferromagnetic switching is illustrated in Fig.~1A which shows the crystal and magnetic structure of CuMnAs, a prototypical antiferromagnetic compound with a high N\'eel temperature\cite{Wadley2013}.  The Mn spin-sublattices with opposite magnetic moments (thick red/purple arrows in Fig.~1A) occupy non-centrosymmetric crystal sites that are inversion partners. The local symmetry properties of the lattice together with the spin-orbit coupling imply that a global electrical current (black arrow) driven through the crystal generates local, oppositely-oriented carrier spin-polarizations (thin red/purple arrows) at the inversion partner sites, with the polarization axis perpendicular to the applied current\cite{Zelezny2014,Wadley2016,Zelezny2017}. This alternating non-equilibrium polarization acts as an effective staggered magnetic field on the antiferromagnetic moments. The strength of the staggered field is proportional to the current-induced polarization and to the exchange coupling between the carrier spins and the antiferromagnetic moments\cite{Zelezny2014,Wadley2016,Zelezny2017}. The physics is analogous to the highly efficient spin-orbit torque switching mechanism in ferromagnets\cite{Chernyshov2009,Miron2011b,Liu2012} whose writing speed is, however, limited by the GHz threshold\cite{Bedau2010,Garello2014}.

To establish the feasibility in antiferromagnets to extend the writing speeds to the THz band, we compare our ultra-short writing pulse experiments  to the results obtained with longer writing pulses in the same device structures. We first recall the previously established technique that has enabled the scaling of the writing pulse time $\tau_p$ down to 250~ps, corresponding to the writing speed $1/\tau_p$ of up to 4~GHz\cite{Wadley2013,Olejnik2017}.  In this scheme, the current pulses are delivered via wire-bonded contacts, and here $\tau_p\sim100$~ps is at the limit achievable with common current-pulse setups. The reversible switching is realized in cross-shape bit cells by alternating current pulses along one of the two orthogonal directions, as illustrated by white dashed lines on the electron microscopy image of the cell in Fig.~1B. White double-arrows depict the corresponding orthogonal N\'eel vector orientations preferably set by the two current directions. Apart from the reversible switching controlled by alternating the two orthogonal writing currents, earlier studies have also shown that multiple-pulses can be applied successively along one writing path, revealing a neuron-like multi-level bit-cell characteristics typical of antiferromagnets\cite{Wadley2016,Olejnik2017,Kriegner2016,Fukami2016}. This has been associated with multi-domain reconfigurations\cite{Grzybowski2016} and we will exploit the feature also in our ps-pulse experiments described below. 

For our experiments with $1/\tau_p$ in the THz range, we use the same cross-shape bit cell and an analogous experimental setup. However, as illustrated in Fig.~1D, we employ a non-contact technique for generating the ultra-short current pulses in the memory cell to overcome the above limit of  common current-pulse setups. To explore reversible writing with  pulses of $\tau_p\approx1$~ps,  we apply THz electro-magnetic transients whose linear polarization can be chosen along two orthogonal directions, as sketched in Fig.~1D. The real wave-form of the electric-field transient is plotted in Fig.~1E \cite{Sajadi2015}.  As in the contact writing scheme, white double-arrows depict the corresponding orthogonal N\'eel vector orientations preferably set by the two orthogonal polarizations.  

In both contact and non-contact setups, the electrical readout is performed by detecting the antiferromagnetic transverse anisotropic magnetoresistance (AMR), as illustrated in Fig.~2\cite{Zelezny2014,Wadley2016,Kriegner2016}. Here the readout current direction is depicted by the white dashed line.  The transverse AMR has an opposite sign for the two orthogonal N\'eel vector directions. To confirm the AMR symmetry, we employed two detection geometries in which we interchanged the readout current and transverse-voltage axes (see Fig.~2). For the AMR, the readout signal flips sign  between the two geometries\cite{Wadley2016}. Note that the AMR readout signal does not depend on the polarity of the switching current. In the present experiments we applied a bipolar wave-form in the contact, larger pulse-time experiments to explicitly highlight the correspondence to the non-contact, ps-pulse measurements (cf. Figs.~1C,E). All experiments are performed at room temperature. 

{\bf Experimental results.} In Fig.~2A we first present measured data for $\mu$s writing pulses delivered by the contact method in a CuMnAs/GaAs bit cell. The pulse-train of one current-direction is turned on for 30~s, then the train is turned off for 30~s followed by turning on for 30~s the pulse-train with the orthogonal current direction. The data show the phenomenology attributed in the earlier studies to the multi-level switching of the antiferromagnet by the current-induced staggered spin-orbit field\cite{Wadley2016,Olejnik2017}. The readout signal increases as the successive pulses within a train arrive at the bit-cell. When the pulse-train is turned off, the readout signal of the memory cell  partially relaxes. (Note that the retention stability can be broadly varied by changing the antiferromagnetic structure parameters\cite{Olejnik2017}.) The sign of the signal is switched when applying the pulse-train with the orthogonal current direction and the overall sign of these reversible switching traces flips between the two readout geometries, consistent with the AMR symmetry. Data in  Fig.~2A were obtained for  an applied writing current density $j=3\times10^7$~Acm$^{-2}$ and a writing pulse repetition rate of 1~Hz. Earlier systematic studies of the multi-level switching in these antiferromagnetic memory devices showed that the magnitude of the signal depends, apart from the number of pulses, on the magnitude of the writing current density and the pulse time\cite{Wadley2016,Olejnik2017}.  

Remarkably, analogous reversible switching traces, with an initial steep increase of the AMR signal followed by a tendency to saturate, can be written in the same  CuMnAs/GaAs memory cell structure by the ps-pulses, as shown in Fig.~2B. Here the writing pulse repetition rate was set to 1~kHz and the current density, recalculated from the applied intensity of the THz electric-field $E=1.1\times 10^5$~Vcm$^{-1}$, was increased  to $j\approx2.7\times10^9$~Acm$^{-2}$ for these ultra-short pulses. (The  $E$ to $j$ conversion is discussed in detail below.) The correspondence between the measured data in Figs.~2A,B indicates that for the ps-pulses the antiferromagnetic switching is also due to the current-induced staggered spin-orbit field. Note that this  switching mechanism allows us to employ the electric-field transient  and that we do not rely on the weak magnetic-field component of the radiation,\cite{Kampfrath2010} or on non-linear orbital-transition effects\cite{Baierl2016}.

To resolve the change in the AMR readout signal induced by a single ps-pulse, we increased the current density to $j\approx2.9\times10^9$~Acm$^{-2}$ and reduced the repetition rate of the writing pulses to 125~Hz and matched it closely to the readout repetition rate (100~Hz). The measured data plotted as a function of the pulse number are shown in Fig.~3A. We observe that the initial ps-pulse accounts for a sizable portion of the total signal generated by the pulse-train. (Note that the scatter in the measured data is likely of an instrumental origin due to the electrical noise from the laser setup.) In Fig.~3B we show corresponding measurements with the $\mu$s-pulses which again highlight the analogous phenomenology of the THz-speed writing and the slower writing in the multi-level antiferromagnetic bit cell.

The electrical current density generated in our cells for a given amplitude of the ps electric-field  pulse (see Fig.~1E) could not be directly measured. In the simplest approach, one can, according to the Ohm's law, divide the THz field by the CuMnAs resistivity. This would result in a device-geometry independent  current density  of $j=8\times10^7$~Acm$^{-2}$ per peak incident THz field of $10^5$~Vcm$^{-1}$. (Here we accounted for the factor of 10 reduction of the THz field in the material due to reflection.) However, when plotting in Fig.~4A the THz-induced switching signal as a function of the incident THz peak field, we find different dependencies in three devices with the width of the central cross of 1, 2, and 3 $\mu$m, and with otherwise the same geometry of the Au-contact pads and approximately the same resistance of the three devices.

We ascribe this observation to the Au electrodes which strongly modify the incident THz field in the cross region. As shown previously,\cite{Novitsky2012,Mcmahon2011} a THz field polarized parallel to the electrode drives currents inside the metal, thereby inducing charges of opposite sign on opposite electrode apexes. These charges, and the resulting voltage across the inner device,  govern the current in our CuMnAs crosses. At a given incident THz peak field, the current flowing through  the three crosses is comparable and, correspondingly,  the current density scales up with decreasing width of the crosses. This is consistent with our observation of increasing AMR signal with decreasing cross size (see main plot of Fig. 4A). Indeed, when we accordingly rescale the data of the main plot of Fig. 4A by the cross sizes, as shown in the inset of Fig. 4A, the three curves fall on top of each other, in agreement with the expected phenomenology for the current-induced spin-torque switching.

We point out that charges at the electrode apexes are induced also when a voltage is applied between  opposite electrodes in the contact experiments. Therefore, similar field distributions in the cross region are expected for the contact (Fig.�1B) and non-contact (Fig. 1C) field applications, differing only by a global scaling factor. We take advantage of this notion to quantitatively calibrate the THz current.  We determine in the contact experiments the critical absorbed Joule energy density leading to device damage for $1/\tau_p$ up to 4 GHz. Since the energy already saturates in the GHz range, extrapolation to the THz writing speed  is straightforward and enables determination of the scaling factor between the incident peak field and the resulting peak current density in the CuMnAs cross.

The calibration procedure allows us to plot the characteristic Joule energy density $\epsilon$ required to obtain a reference switching signal of, e.g., 1~m$\Omega$ up to the THz range, as shown in the inset of Fig. 4B. We find a steeply decreasing $\epsilon$ with increasing  $1/\tau_p$ below $\sim$MHz to a saturated value of $\epsilon$ in the GHz range (see also main plot of Fig.~4B) that extends to the THz writing speed. This  demonstrates that current-induced switching at the THz writing speed is as energy-efficient as at the GHz speed. 

The Joule energy leading to device damage and the energy required for the reference 1~m$\Omega$ switching signal are separated by approximately a factor of 2  in both the GHz and THz writing-speed range. This breakdown margin is sufficiently large, allowing us to demonstrate tens of thousands of reversible write-read cycles without any notable device wear-out\cite{Olejnik2017}.

In Fig.~4C we finally demonstrate  that one antiferromagnetic multi-level bit-cell can be simultaneously addressed by the non-contact THz-speed writing and the contact lower-speed writing. For the illustration we choose a CuMnAs/Si bit cell and plot the readout signal when applying 200~ms pulses of $j=1\times10^7$~Acm$^{-2}$ by the contact method with alternating orthogonal current-path directions and a delay between pulses of 5~s. After 15 switchings we added the non-contact writing in the form of a 90~s long train of ps-pulses with a kHz repetition rate and  a current density $j\approx 1.6\times10^9$~Acm$^{-2}$. The non-contact ps-pulses  induce an additional switching of the multi-level cell with the superimposed smaller switching signals controlled by the contact pulses.

{\bf Discussion.} We will now discuss our results in the broader context of writing of magnetic memories. Data presented in Figs.~2-4 illustrate that we have pushed the electrical spin-torque switching, that drives current research and development of magnetic memories, to the THz writing-speed range. 
At the GHz writing speed, the antiferromagnetic CuMnAs bit cells are written by current densities $j\sim10^8$~Acm$^{-2}$\cite{Olejnik2017}. These are the same values as reported in ferromagnetic spin-orbit torque MRAMs\cite{Garello2014,Prenat2016}. For the THz writing speeds, the applied current density in our antiferromagnetic devices increases only to  a value of the order of $j\sim10^9$~Acm$^{-2}$ and the writing  energy remains comparable to that of the GHz writing speed. This makes antiferromagnets realistic candidate materials for electrically controlled memory devices in the THz band. 

The result is in striking contrast to ferromagnets where the projected current and energy of the writing would increase by three orders of magnitude from the GHz to the THz writing speed.
For  $1/\tau_p$ up to a GHz, the current  density required for switching in spin-orbit  torque memories fabricated in the common ferromagnetic transition-metal structures is $\sim10^8$~Acm$^{-2}$ and only weakly deviates from the steady-current limit, owing to heat-assisted magnetization reversal\cite{Garello2014}. This current density corresponds to an effective field $H_{eff}\sim 10-100$~mT required for the switching of ferromagnetic moments  over the magnetic energy barrier. However, when  $1/\tau_p$ is above the GHz threshold, the steady-current limit $H_{eff}$ is no longer sufficient for switching because $\tau_p$ becomes comparable or smaller than the limiting magnetization reorientation time-scale. This is given by $1/f$ where $f=\frac{\gamma}{2\pi}H_{eff}$ is the ferromagnetic resonance frequency. To keep $f$ in scale with $1/\tau_p$, the effective writing field, and therefore $j$, have to increase linearly with $1/\tau_p$ above the GHz threshold\cite{Garello2014}. In particular, $f$ reaches 1~THz at about 30~T and the switching current density linearly extrapolated to the THz writing speed would be $\sim10^{11}$~Acm$^{-2}$ in the ferromagnetic spin-orbit torque devices. This also means that while the writing energy $\sim j^2\tau_p$ initially drops down with increasing  $1/\tau_p$, in ferromagnets it starts to increase linearly with $1/\tau_p$ above the GHz threshold. A THz writing speed would then require three orders of magnitude higher energy than the GHz writing speed used in present MRAM devices. With these basic physical limitations, the current-induced spin-orbit torque switching has not been pushed in ferromagnetic MRAMs to $1/\tau_p$ far exceeding 5~GHz\cite{Garello2014,Prenat2016}. 

On the other hand, the antiferromagnetic resonance frequency is exchange-enhanced and scales as $\sim\sqrt{H_E H_A}$, which puts it in the THz range\cite{Kittel1951}. Here $H_E$ is the inter-spin-sublattice exchange field and $H_A$ stands for the anisotropy field (in the absence of externally applied fields). This enhancement of the antiferromagnetic resonance frequency and, correspondingly, of the threshold writing speed is due to canting of the antiferromagnetic spin-sublattices when brought out of equilibrium. 

A uniform magnetic field required for switching is also exchange-enhanced in antiferromagnets and scales as  $\sim\sqrt{H_EH_A}$ which typically takes it into the tens of Tesla range, inaccessible in microelectronics. However, we exploited the current-induced staggered field that is commensurate with the staggered N\'eel order in  the antiferromagnet, for which the exchange-enhancement factor of the switching field amplitude is absent\cite{Zelezny2014}. This  principle allows antiferromagnets to reach THz writing speeds at accessible fields and energies, as confirmed by our experiments and predicted in earlier numerical simulations\cite{Zelezny2014,Roy2016}. 

We note that in our present experiments, electrical readout is performed at macroscopic time-delay scales compared to the ps-scale of the writing pulses, taking advantage of the memory effect in our antiferromagnetic CuMnAs bit cells. Among our envisaged future research directions is to combine  the ps-scale writing with a comparable readout scale. This will provide a time-resolved physical picture of the switching mechanism  and will open the prospect of an ultra-fast complete write-read cycle in antiferromagnetic memories. In  an earlier work by several of us\cite{Saidl2017} we have already demonstrated an initial step in this direction by detecting the N\'eel vector direction in CuMnAs in a fs-laser setup by magnetic linear dichroism, which is the optical counterpart of the AMR. 

As a concluding remark we point out that a THz-speed memory, whether realized in antiferromagnets or another alternative system that may be discovered in the future, is only one of the many components that need to be developed to make true THz electronics and information technologies a realistic prospect. In the meantime, however, the ultra-fast writing of antiferromagnetic bit cells can be potentially exploitable without separate THz-speed processors. The multi-level neuron-like characteristics allows for integrating memory and logic within the antiferromagnetic bit cell, as illustrated in earlier measurements on the example of a pulse-counter functionality\cite{Olejnik2017}.  Future low-noise experiments with THz  writing pulses and repetition rates spanning a broad range up to THz will establish the feasibility and versatility of this autonomous THz memory-logic concept built within the antiferromagnetic bit-cells, that requires no separate processor to perform the THz-speed logic operation.

\bigskip
\noindent{\large\bf Acknowledgements}

\noindent We acknowledge support from the Ministry of Education of the Czech Republic Grants LM2015087 and LNSM-LNSpin, the Grant Agency of the Czech Republic Grant No. 14-37427, the University of Nottingham EPSRC Impact Acceleration Account grant No. EP/K503800/1, the Alexander von Humboldt Foundation, the ERC Synergy Grant SC2 (No. 610115), the Transregional Collaborative Research Center (SFB/TRR) 173 SPIN+X, the ERC Consolidator Grant TERAMAG (No. 681917), and the Swiss National Science Foundation Grant (No. 21-63864).

\protect\newpage

\begin{figure}[h!]
\hspace*{-0cm}\epsfig{width=1\columnwidth,angle=0,file=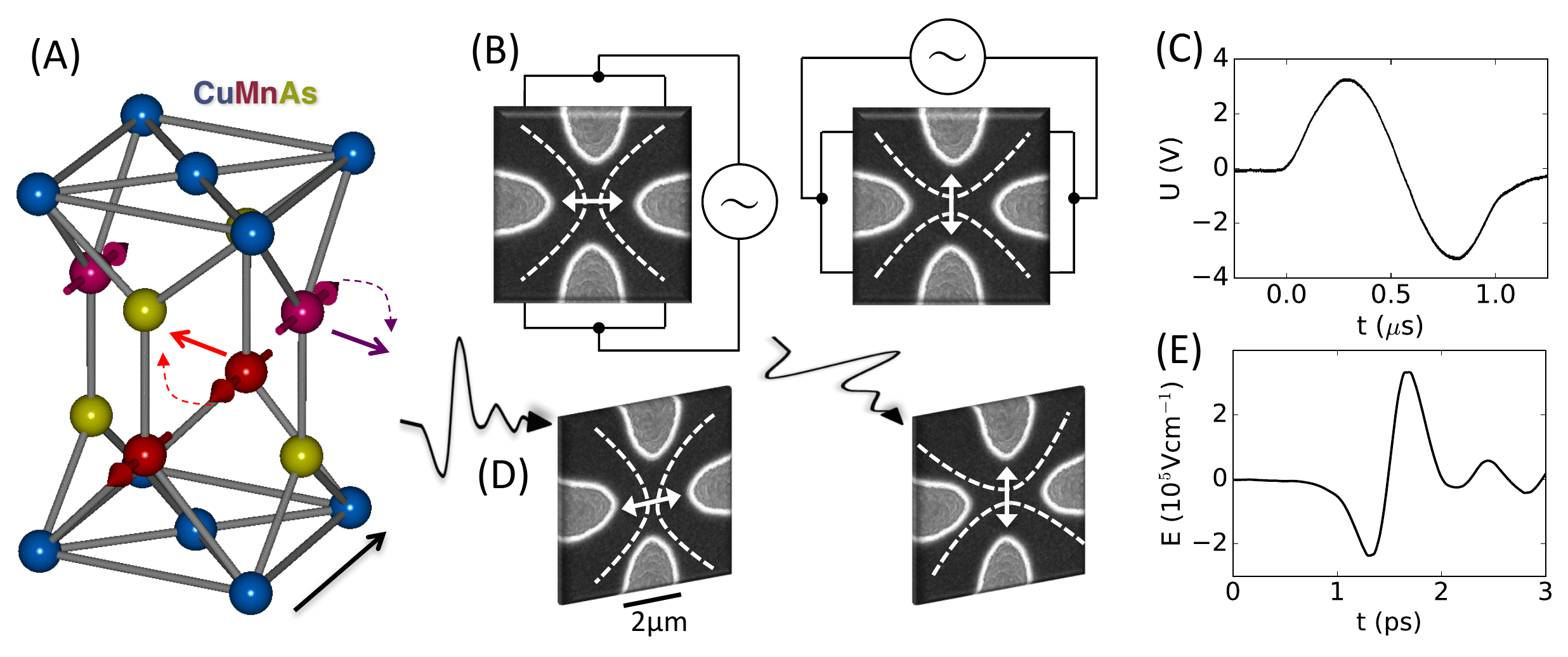}
\caption{(A) Schematics of the crystal and magnetic structure of the CuMnAs antiferromagnet. An electrical current (black arrow) generates a staggered spin-orbit field (thin red/purple arrows) that switches the antiferromagnetic moments (thick red/purple arrows). (B) Electron microscopy image of the cross-shape bit cell and schematics of the reversible writing by electrical pulses of two orthogonal current directions delivered via wire-bonded contacts.  White dashed lines illustrate electrical current paths and white double-arrows the corresponding preferred N\'eel vector orientations. (C) Wave-form of the applied $\mu$s  electrical pulses.  (D) Schematics of the reversible writing by  THz electric-field transients whose linear polarization can be chosen along two orthogonal directions. (E) Wave-form of the applied ps radiation pulses.}
\label{fig1}
\end{figure}

\begin{figure}[h!]
\hspace*{-0cm}\epsfig{width=1\columnwidth,angle=0,file=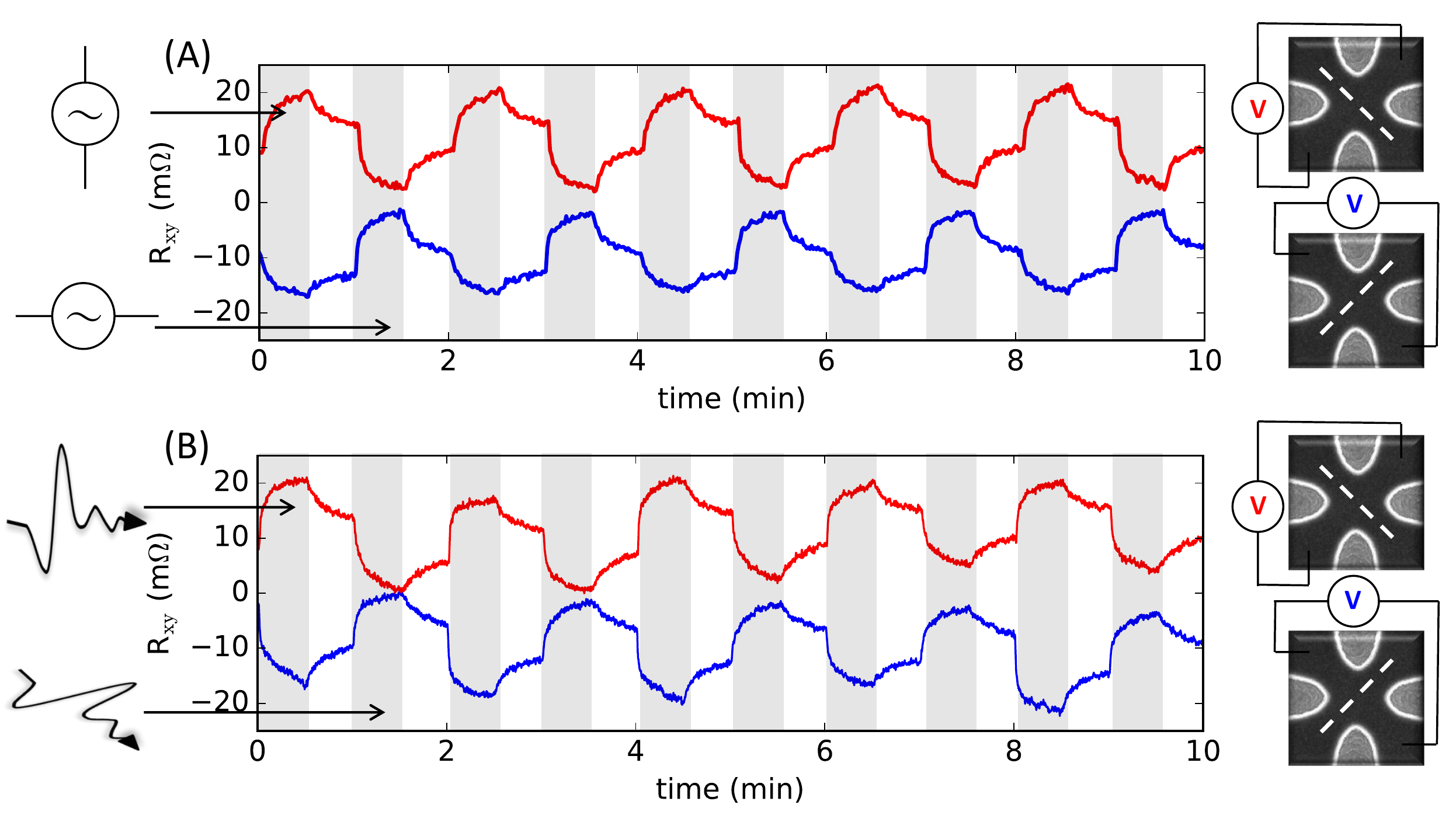}
\caption{(A) Reversible multi-level switching by 30~s trains of  $\mu$s electrical pulses with a Hz pulse-repetition rate, delivered via wire-bonded contacts along two orthogonal directions. The applied writing current density in the 3.5~$\mu$m-size CuMnAs/GaAs cell is $3\times10^7$~Acm$^{-2}$. Intervals with the  pulse trains turned on are highlighted in grey and the two orthogonal current-directions of the trains are alternating from one interval to the next. Electrical readout is performed at a 1~Hz rate. Right insets show schematics of the transverse AMR readout.  White dashed lines depict readout current paths. (B) Same as (A) for ps-pules with a kHz pulse-repetition rate.  The writing current density in the 2~$\mu$m-size CuMnAs/GaAs bit cell recalculated from the amplitude of the applied THz electric-field transient  is   $2.7\times10^9$~Acm$^{-2}$.  Electrical readout is performed at a 8~Hz rate.}
\label{fig2}
\end{figure}

\begin{figure}[h!]
\hspace*{-0cm}\epsfig{width=1\columnwidth,angle=0,file=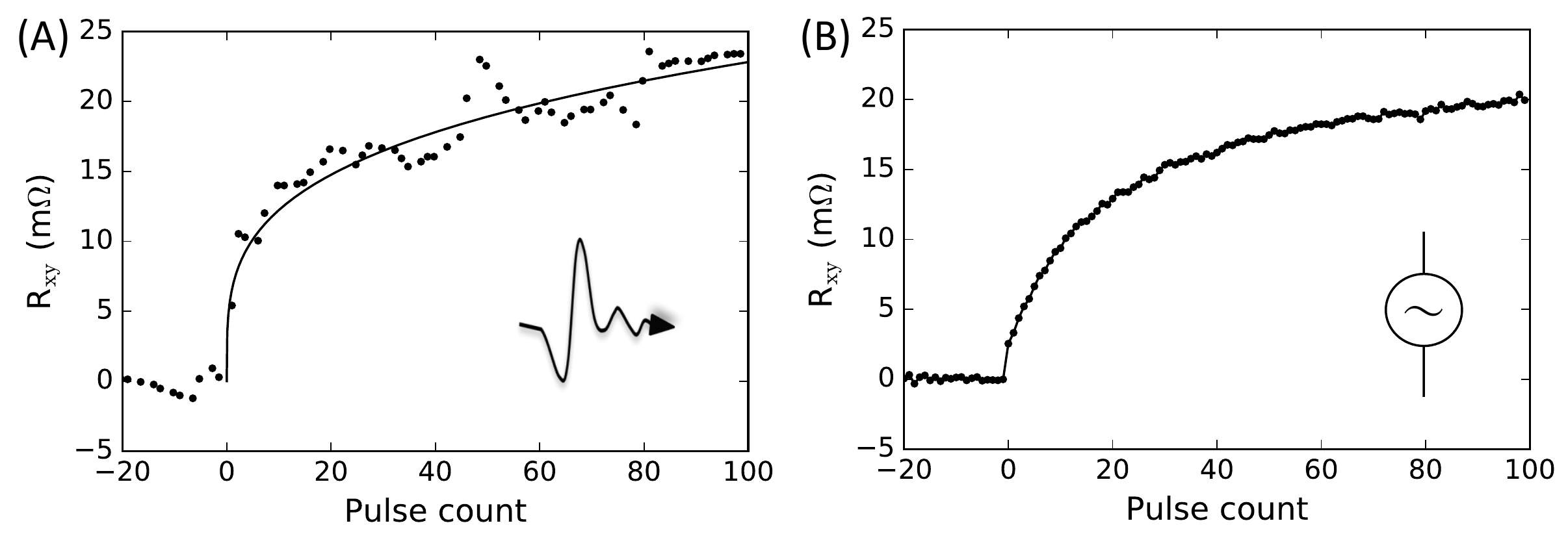}
\caption{(A) The multi-level memory signal as a function of the number of applied ps-pulses. The writing current density in the 2~$\mu$m-size CuMnAs/GaAs bit cell recalculated from the applied THz field amplitude is   $2.9\times10^9$~Acm$^{-2}$. (B) Same as (A) for the $\mu$s-pulses and an applied writing current density of $3\times10^7$~Acm$^{-2}$ in the 3.5~$\mu$m-size CuMnAs/GaAs cell.}
\label{fig3}
\end{figure}

\begin{figure}[h!]
\vspace*{-1cm}
\hspace*{-0cm}\epsfig{width=1\columnwidth,angle=0,file=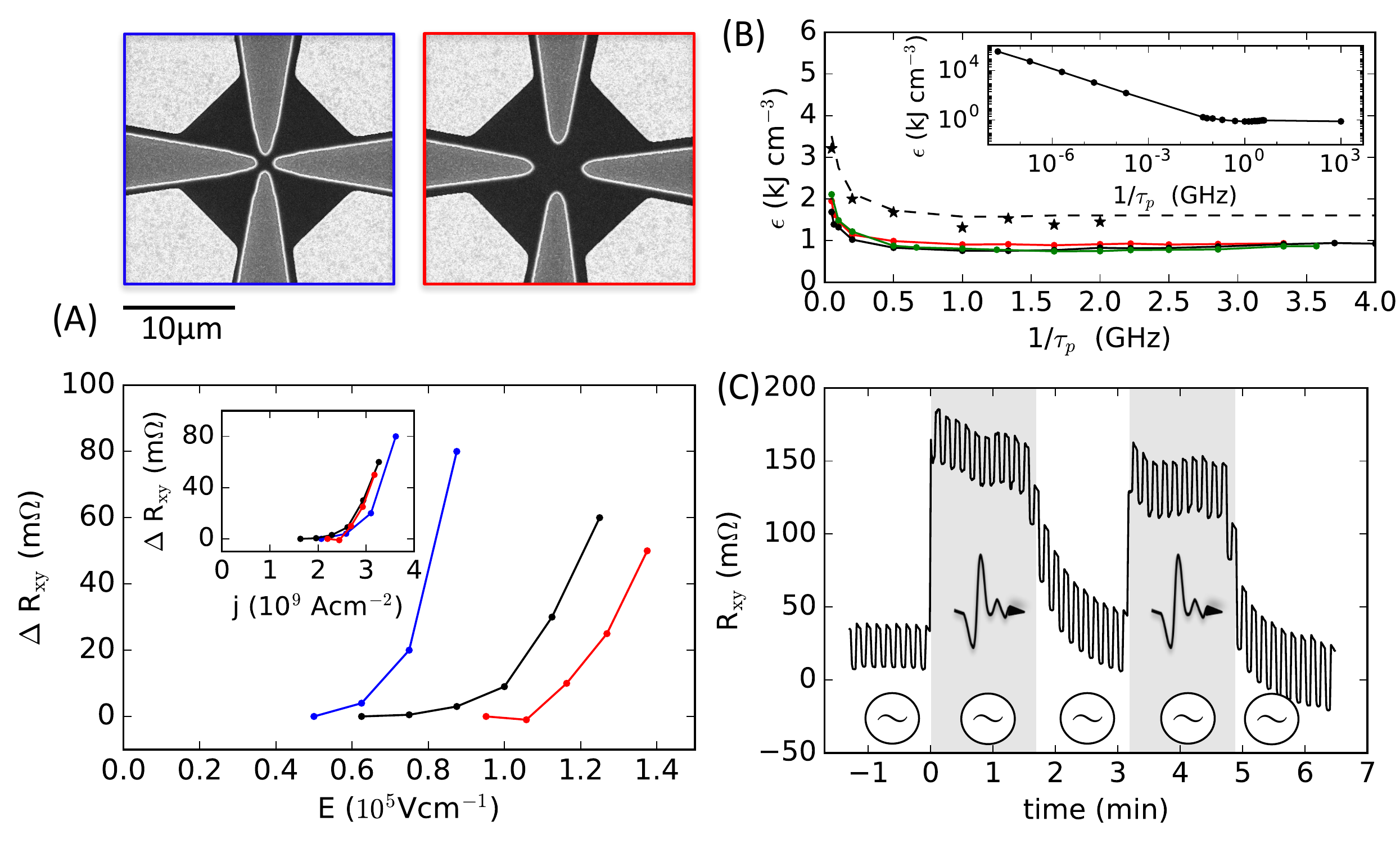}

\caption{(A) Magnitude of the switching signal as a function of the THz field amplitude (main panel) and of the converted current density (inset) for 1, 2 and 3~$\mu$m size devices (blue, black, and red dots). Top insets show electron microscopy images of the  1 and 3~$\mu$m size devices. Light regions are Au-contact pads, grey regions are etched down to the GaAs substrate, and black regions are CuMnAs. (B) Writing energy density (black, red, and green dots), $\epsilon=\rho j^2\tau_p$,  required to obtain a 1~m$\Omega$ switching signal as a function of the writing speed $1/\tau_p$ in the linear scale (main plot) and in the log-log scale (inset). All data points, except for the point at $1/\tau_p$=1~THz in the inset are obtained from the contact measurements. The point at $1/\tau_p$=1~THz in the inset is from the non-contact measurement using the $E$ to $j$ conversion based on the breakdown energy (see text).   Black dots in the main plot correspond to 2~$\mu$m,  red to 3~$\mu$m, and  green to 4~$\mu$m  size CuMnAs/GaAs bit cells. Black star-symbols and dashed line represent the limiting breakdown energy density. (C) Contact writing by 200~ms pulses of current density $1\times10^7$~Acm$^{-2}$ (white intervals) and the contact writing superimposed on the non-contact writing by a train of ps-pulses with a kHz repetition rate and a THz field amplitude corresponding to an additional current density of  approximately $1.6\times10^9$~Acm$^{-2}$ (grey intervals). Data were measured in a 10~$\mu$m-size CuMnAs/Si bit cell.}
\label{fig4}
\end{figure}
\end{document}